\documentclass[prl,twocolumn,superscriptaddress,nofootinbib]{revtex4-1}
\usepackage{graphicx}
\usepackage{bm}
\usepackage{color}
\usepackage[colorlinks,citecolor=blue,urlcolor=blue,linkcolor=blue]{hyperref}

\begin{document}
	
	\title{Exploring Supersymmetry with machine learning}
	
	\author{Jie Ren}
	
	\affiliation{Department of Physics and Institute of Theoretical Physics, Nanjing Normal University, Nanjing, 210023, China}	
	\affiliation{CAS Key Laboratory of Theoretical Physics, Institute of Theoretical Physics, Chinese Academy of Sciences, Beijing 100190, China}
	\affiliation{School of Physics, University of Chinese Academy of Sciences, Beijing 100049, China}
	
	\author{Lei Wu}
	
	\affiliation{Department of Physics and Institute of Theoretical Physics, Nanjing Normal University, Nanjing, 210023, China}
	
	\author{Jin Min Yang}
	
	\affiliation{CAS Key Laboratory of Theoretical Physics, Institute of Theoretical Physics, Chinese Academy of Sciences, Beijing 100190, China}
	\affiliation{School of Physics, University of Chinese Academy of Sciences, Beijing 100049, China}
    \affiliation{Department of Physics, Tohoku University, Sendai 980-8578, Japan}

	\author{Jun Zhao}
	
	\affiliation{CAS Key Laboratory of Theoretical Physics, Institute of Theoretical Physics, Chinese Academy of Sciences, Beijing 100190, China}
	\affiliation{School of Physics, University of Chinese Academy of Sciences, Beijing 100049, China}
	
	\begin{abstract}
Investigation of well-motivated parameter space in the theories of Beyond the Standard Model (BSM) plays an important role in new physics discoveries. However, a large-scale exploration of models with multi-parameter or equivalent solutions with a finite separation, such as supersymmetric models, is typically a time-consuming and challenging task. In this paper, we propose a self-exploration method, named Machine Learning Scan (MLS), to achieve an efficient test of models. As a proof-of-concept, we apply MLS to investigate the subspace of MSSM and CMSSM and find that such a method can reduce the computational cost and may be helpful for accelerating the exploration of supersymmetry.
	\end{abstract}
	
	\maketitle
	
	
	\textit{Introduction.}
	The discovery of 125 GeV Higgs boson~\cite{Aad:2012tfa, Chatrchyan:2012xdj} completes the Standard Model (SM) and confirms the mass generation mechanism through the spontaneous electroweak symmetry breaking. However, there are still some longstanding problems in particle physics, such as the hierarchy problem and dark matter (DM), which motivates various extensions of the SM. Up to now, the vast theoretical and experimental efforts have been devoted to searching for new physics.
	
	In order to know the status and prospects for new physics models, one often needs to scrutinize the parameter space under all available experimental constraints. Several modern computing methods, for example, MCMC~\cite{MacKay:2002:ITI:971143} and MultiNest~\cite{Feroz:2007kg, Feroz:2008xx}, have been used to accelerate such numerical analyses of the new physics models by high energy physics (HEP) packages, e.g.~\cite{Lafaye:2004cn, Bechtle:2004pc, deAustri:2006jwj, Allanach:2007qj, Feroz:2008xx, Strege:2014ija, Han:2016gvr, Athron:2017ard, Bagnaschi:2017tru}. Even though, those calculations are usually time-consuming and challenging, in particular for models with multi-parameter, multi-modal or large, curving degenerate space.
	
	
	Machine learning (ML) is becoming a powerful tool for the study of particle physics, since it can efficiently find the patterns hidden in complex and large data sets, and then apply them to unseen samples. A great successful example of this is the use of boosted decision trees \cite{LiorRokach2007} in the LHC experiment that led to the Higgs discovery. Recently, more and more advanced ML techniques, such as deep neural network (NN) \cite{Bechtle:2017vyu}, have been applied to the studies of BSM phenomenology \cite{Bridges:2010de, Buckley:2011kc, Bornhauser:2013aya, Caron:2016hib, Bertone:2016mdy}.
	
	Several existing ML works are managed to accelerate the analysis of new physics models~\cite{Bridges:2010de, Buckley:2011kc, Bornhauser:2013aya, Caron:2016hib, Bertone:2016mdy, Bechtle:2017vyu}, however, they adopted the offline-trained ML models to approximate physical quantities. The performance strongly depends on the complexity of used ML models and the quality of initial training samples, which is usually difficult to reach high local accuracy. On the other hand, an online-training package BAMBI~\cite{Graff:2011gv} allows a shallow neural network to evolve over time. But the use of NN in BAMBI is merely to improve the local likelihood evaluation in the MultiNest and is impossible to discover new separate and degenerate target regions globally.
	
	In this paper, we propose a new numerical method of analyzing models, namely Machine Learning Scan (MLS). It can iteratively self-explore the parameter space from scratch by guiding the sampling from the incrementally learned knowledge. The efficiency of sampling is greatly increased by the active learning approach and the ML models are constantly optimized by the incremental learning method. The accuracy of learned physical observables are improved by deep learning techniques. All collected samples, together with extra random points added, are used to recover the parameter space. We apply our MLS to four benchmark models and compare MLS with other existing methods. We find that the MLS can reduce the computational cost and ensure a better discovery of target regions.
	
	
	\textit{Method.}
	The likelihood test is widely used to measure the quality of a parameter point $\bm{x}$ against experimental data,
	\begin{equation}
	\mathcal{L}(\bm{x}) = \prod_i \mathcal{L}_i(O_i(\bm{x}); O^*_i, \sigma^*_i) ,
	\end{equation}
	where $O_i$ are the theoretical predictions given by a certain model, and $O^*_i$ are the corresponding experimental data with uncertainty $\sigma^*_i$. Here the correlation of experimental data is ignored for simplicity. In models with multi-parameter, $\mathcal{L}(\bm{x})$ often vanishes in most part of parameter space and is non-trivial only in regions of thin sheets or hypersurfaces. Besides, the likelihood function may have equivalent values with a finite separation. So, using limited number of random samples, it is difficult to obtain a well reconstructed likelihood function by training ML models only once. On the other hand, the physical observables usually vary smoothly with parameters and can be easily learned by ML models. Thus, we can directly train ML models to learn the physical observables, rather than the likelihood function, with sufficient accuracy in our calculations, e.g.,
	\begin{equation}
	{O}_i(\bm{x}) \approx \hat{O}_i(\bm{x}) = M_i(\bm{x}) ,
	\end{equation}
	where $\hat{O}_i$ is the approximation of the physical observables $O_i$, which is given by the ML model $M_i$. Without calling HEP packages, such approximate calculations can be very fast. Thanks to the powerful generalization of ML models, the likelihood function can be then indirectly reconstructed as
	\begin{equation}
	\hat{\mathcal{L}}(\bm{x}) = \prod_i \mathcal{L}_i(\hat{O}_i(\bm{x}); O^*_i, \sigma^*_i) .
	\end{equation}
	
	\begin{figure}
		\includegraphics[width=8.5cm]{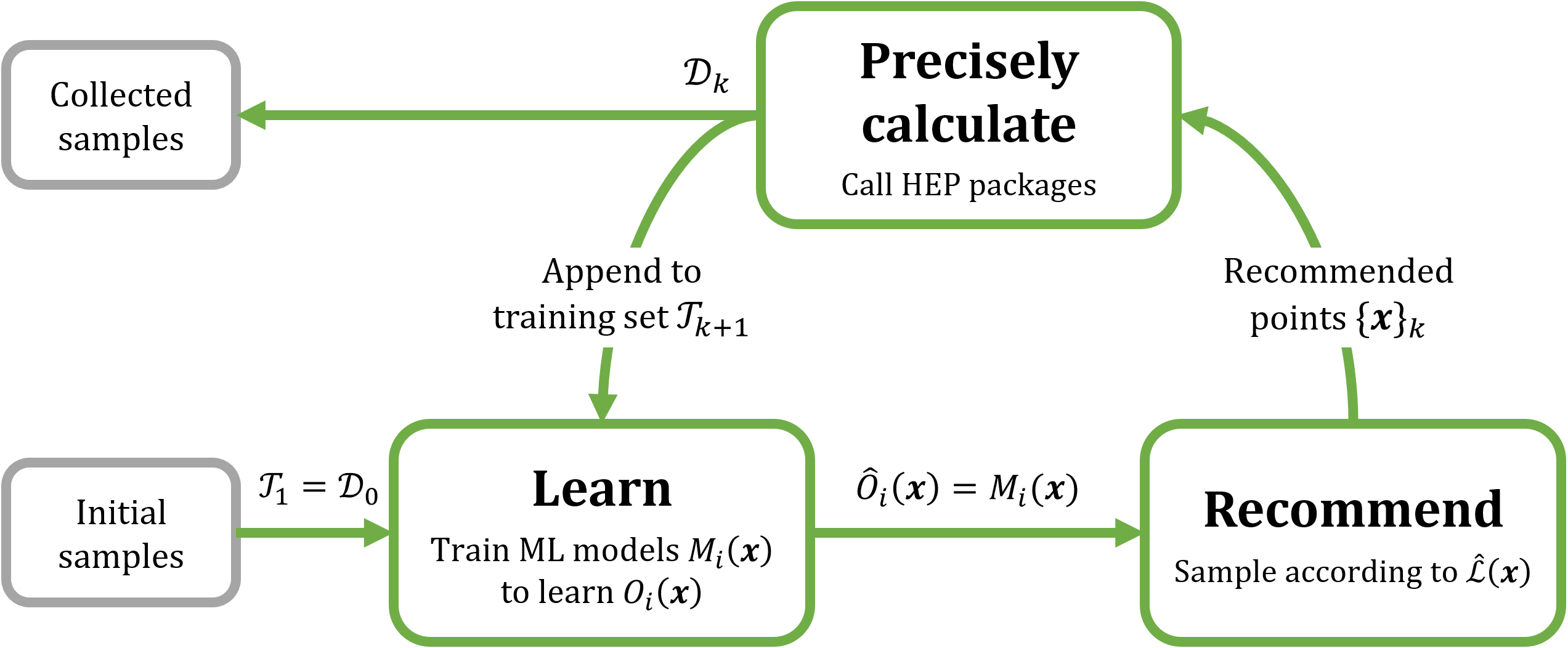}
		\caption{The MLS works iteratively. First, train machine learning models using the already collected samples as training data. Then, sample the important regions according to the reconstructed likelihood. Next, calculate observables of the recommended points using HEP packages, and append these samples to the training set to improve the machine learning models in the next iteration. The procedure repeats until sufficient target samples are collected.}
		\label{workflow}
	\end{figure}
	
	Applying ML approaches to evaluating parameter space requires sufficient amount of data to train ML models at the beginning, which can be a heavy overhead. Therefore, we adopt the policy of active learning and incremental learning to improve this disadvantage.
	
	The overall workflow of our method is schematically shown in Fig.~\ref{workflow}. In each iteration $k$, the approximate function of physical observables $\hat{O}_i$ are learned from all the collected samples $\mathcal{T}_k$. Then the machine actively finds and samples the important regions, according to the reconstructed likelihood $\hat{\mathcal{L}}$, to recommends some important parameter points, and precisely calculates their corresponding physical observables with the help of HEP packages. These are used as the feedback $\mathcal{D}_k$ and are appended to the training set $\mathcal{T}_{k+1} = \mathcal{T}_k \cup \mathcal{D}_k$ to incrementally improve the $\hat{O}_i$ in the next iteration $k+1$. The process repeats until sufficient target samples are collected. In such a way, the ML models are gradually improved by more collected samples, and the important regions are in turn continually refined due to the increasing accuracy of the ML models. As a result, samples accumulate around the target regions. The ML models become very accurate in the target regions and work well in the whole parameter space. The sampling efficiency is therefore greatly increased because seldom points are sampled in the unimportant regions. The key details of our implementation in MLS are the followings:
	
	\begin{itemize}
		\item The initial small batch of samples $\mathcal{T}_0$ are usually randomly generated, and previously existing samples are worth to be included.
		
		\item The MLS is a general framework. It does not specify the type of ML models. The default ML model is the deep fully connected neural network whose fitting capability is ensured by~\cite{Cybenko1989, HORNIK1991251}, but any ML model with sufficient representational capacity and generalization capability can be used.
		
		\item Given that millions of ML model evaluations can be done in a trivial time, the simple rejection sampling taking $\hat{\mathcal{L}}$ as the target distribution is adopted as the default setting to sample the important regions. Also, human experiences on the parameter space to be explored are encouraged to be applied on the ML models and the sampling method, which may increase the performance further to some extent.
		
		\item Besides the parameter points sampled according to $\hat{\mathcal{L}}$, it is essential to recommend some random points to enhance the exploration. Recommending more random points, we encourage the machine to explore unknown regions; otherwise, it will mainly focus on the known important regions.
		
		\item All the collected samples are fully utilized to train the ML models, without validation, test and regularization, since calculating observables precisely using HEP packages is time-consuming. Training with a subset of samples may lead to the loss of some critical structures, and regularizing the ML models may smooth out some sharp structures of the parameter space. Overfitting is inevitable but controllable. Wrong search directions can be rectified in the following iterations by virtue of the active learning and incremental learning.
	\end{itemize}
	
	It is also worth noting that the resulting samples are accurate, because all the points recommended in each iteration should be fed into the HEP packages to calculate their observables precisely. Moreover, the resulting ML models are accurate enough which can be used for further fast evaluation of the parameter space.
	
	We implement our MLS framework in Python with the open-source deep learning framework PyTorch \cite{pytorch} to build and train ML models. The computations are performed using Intel Core i7-4930K and NVIDIA Titan XP. To demonstrate the ability of our method, we firstly apply it to two toy models: \textit{Model-1} with separate equivalent solutions, and \textit{Model-2} with multi-parameter. Then, we use MLS to study two supersymmetric models \textit{Model-3} and \textit{Model-4}, which are the constrained MSSM and the MSSM with alignment limit, respectively.
	
	
	\textit{Model-1.}
	The eggbox model, which is given by
	\begin{equation}
	\label{model1}
	O_{1}(x_1, x_2) = \left\{ 2 + \cos \frac{x_1}{2} \cos \frac{x_2}{2} \right\}^5 ,
	\end{equation}
	The function $O_{1}$ has an eggbox shape in the parameter space $[0, 10 \pi]^2$. Suppose that the ``experimental'' value is $O^* = 100$ with standard deviation $\sigma^* = 10$ and the likelihood function is Gaussian $\mathcal{L}(x_1, x_2) = \exp \left\{ -\frac{(O(x_1, x_2) - O^*)^2}{2 {\sigma^*}^2} \right\}$. We require our samples to satisfy the ``experimental" observable within $2 \sigma$ region. The target regions are well separated thin annuli with high likelihood values.
	
	\begin{figure}[t]
		\includegraphics[width=8.5cm]{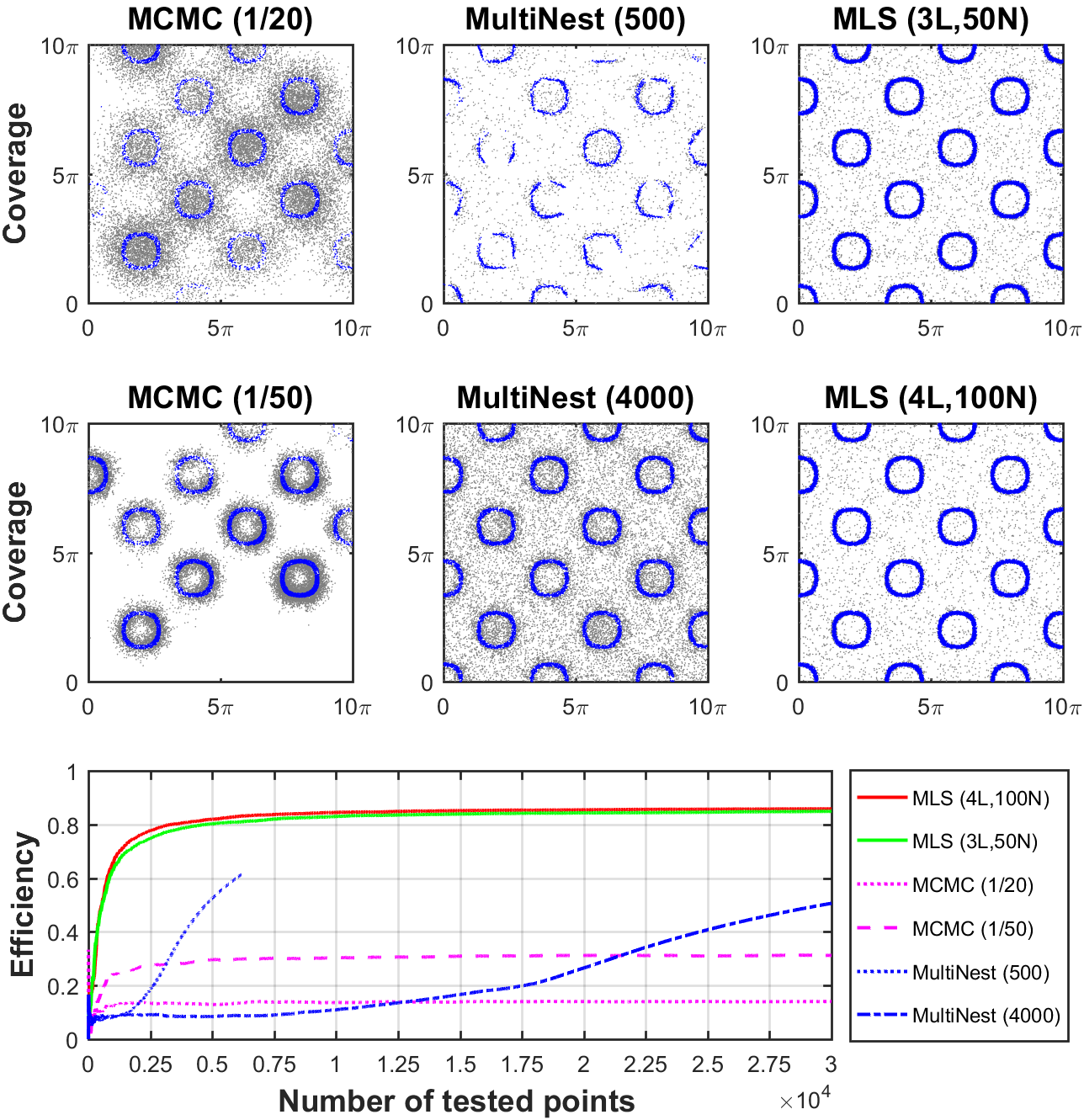}
		\caption{Coverage situation and efficiency of the MLS, MCMC and MultiNest for exploring the eggbox model in Eq.~\ref{model1}, where the number of tested points is $3\times 10^4$ except the case of MultiNest with 500 live points. The algorithmic parameters in brackets are defined in the context. The blue and gray points denote tested points inside and outside the target regions, respectively.}
		\label{toy_eggbox_dist}
	\end{figure}
	
	In Fig.~\ref{toy_eggbox_dist}, we show the coverage situation (the distribution of tested points over target regions) and the efficiency (number of discovered solutions / number of tested points) of the MLS, MCMC and MultiNest for exploring the model defined in Eq.~\ref{model1}. For the MCMC method, a set of 20 chains using the Gaussian proposal with a large step value (1/20 of the parameter range) wander across several target regions, while another set of 20 chains with a small step value (1/50 of the parameter range) are trapped into local target regions. The efficiency is quite low. For the MultiNest method, a small number of live points (500) is difficult to discover all equivalent solutions of the likelihood function. While another run can discover all the solutions with an optimized number of live points (4000), but has a much slowly increasing efficiency. For our MLS method, we construct a 3(4)-hidden-layer deep NN to learn the observable. The hidden neurons, 50(100) neurons per layer, are activated by the rectified linear unit (ReLU) and the output neuron is linear. The standard Adam optimizer~\cite{DBLP:journals/corr/KingmaB14} is adopted to train the NN with the mean-squared-error loss function up to 1000 epochs using a constant learning rate 0.001. With 100 initial random samples, the machine recommends 100 points (including 10 random ones) in each iteration. In general, the MLS is insensitive to algorithmic parameters so that both choices have high efficiency and better coverage to find all the solutions.
	
	\begin{figure}[t]
		\includegraphics[width=8.5cm,height=4cm]{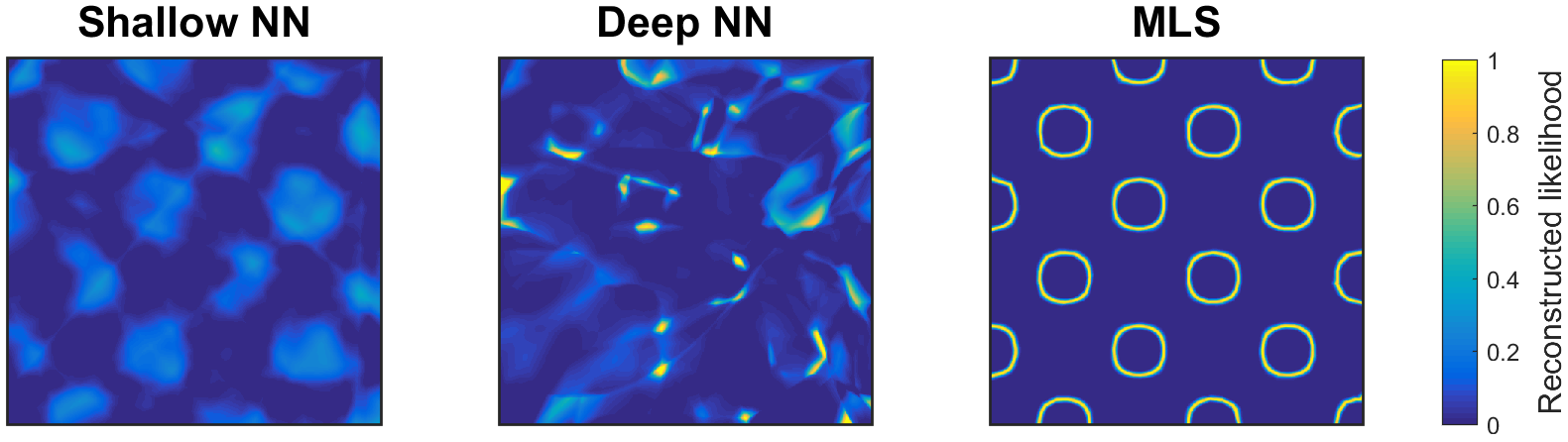}
		\caption{Reconstructed likelihood function of the eggbox model learned by shallow NN, deep NN and our MLS using 2000 samples. The colormap denotes the reconstructed likelihood values.}
		\label{toy_eggbox_fit}
	\end{figure}
	
	Besides, in Fig.~\ref{toy_eggbox_fit}, we also present the reconstructed likelihood functions learned by a shallow NN (single hidden layer with 2000 neurons) and a deep NN (4 hidden layers with each having 100 neurons), respectively. The NNs are trained only once to directly learn the likelihoods of 1800 random samples. The best models are chosen based on the validation on an independent set of 200 random samples. Compared with our MLS with 2000 collected samples, they are much worse to reconstruct the likelihood function. The success of the MLS lies in the fact that it can instruct itself to self-improve its knowledge of the parameter space to better reconstruct the likelihood function than other approaches.
	
	
	\textit{Model-2.}
	The 2D and 7D quadratic model, which is defined by
	\begin{equation}
	\label{model2}
	O(\bm{x}) = \bm{x}^\mathrm{T} \bm{x} = \sum_{i=1}^n x_i^2 ,
	\end{equation}
	where $\bm{x}$ is an $n$ dimensional vector. The likelihood function $\mathcal{L}(\bm{x}) = \exp \left\{ -\frac{(O(\bm{x}) - O^*)^2}{2 {\sigma^*}^2} \right\}$, where the ``experimental'' data is assumed as $O^* = 2$ with the standard deviation $\sigma^{*} = 0.1$. We require our samples to satisfy the ``experimental" observable within $2 \sigma$ region. The target region is a hyperspherical surface in high dimensions. The used NN architecture of MLS is the same as that in \textit{Model-1}. For the 2D (7D) model, the NN is initialized using 20 (100) random samples, and 10 (100) points with 10\% random ones are recommended in each iteration.
	
	\begin{figure}[t]
		\includegraphics[width=8.5cm,height=5cm]{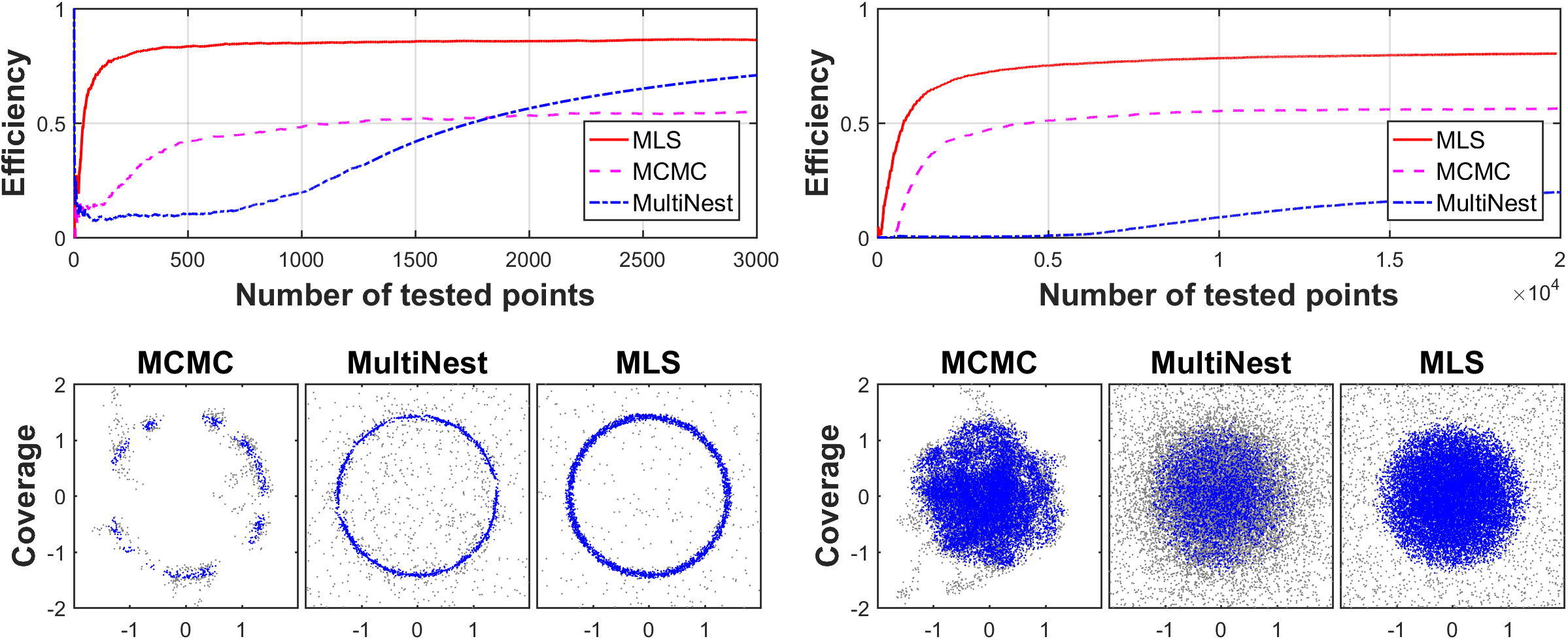}
		\caption{Same as Fig.~\ref{toy_eggbox_dist}, but for 2D (left) and 7D (right) quadratic models in Eq.~\ref{model2}. The total number of tested points is 3000 for 2D and $2 \times 10^4$ for 7D, respectively.}
		\label{toy_quadratic}
	\end{figure}
	
	In Fig.~\ref{toy_quadratic}, we compare the efficiency and coverage situation of the MLS, MCMC and MultiNest for exploring the 2D and 7D models given by Eq.~\ref{model2}. The MCMC and MultiNest methods are optimized for reaching good coverage with least number of tested points. For the MCMC, we use the Gaussian proposal with an optimized step value (1/50 of the parameter range) and run 10 chains. For the MultiNest, the number of live points in 2D and 7D are taken as 200 and 1000, respectively. We can see that our MLS has a much better efficiency and coverage situation than the MCMC and MultiNest for the same amount of tested points. Up to 1000 tested points, the efficiency of MLS decreases from 84.5\% for 2D to 55.5\% for 7D, while the corresponding values of MCMC and MultiNest reduce from 48.2\% to 22.9\%, and from 19.6\% to 0.4\%, respectively. Moreover, the coverage situation of MCMC is the worst, because it is usually trapped in parts of target regions with local maximal likelihood.
	

	\textit{Model-3}.
Among various SUSY models, CMSSM has been widely investigated. We apply the MLS to analyze the CMSSM and compare its results with the MultiNest. Given our limited computing resource, we scan the following parameter space of CMSSM.~\footnote{A comprehensive scan over the whole parameter space of CMSSM will appear soon. The code will be also provided along with the paper in arXiv.}:
	\begin{eqnarray*}
		5~\mathrm{TeV} < M_0 < 10~\mathrm{TeV},  \quad 1~\mathrm{TeV} < M_{1/2} < 10~\mathrm{TeV}, \\
		|A_0| < 10~\mathrm{TeV}, \quad 3 < \tan\beta < 70, \quad \mathrm{sign}(\mu) = -1
	\end{eqnarray*}
	All the SM parameters are fixed in our calculation. To perform a real comparison with a real likelihood function under realistic assumptions, we adopted flat priors and the likelihood function defined in the GAMBIT~\cite{Athron:2017qdc}. The total likelihood function consists of contributions from electroweak precision measurements (with \textsf{PrecisionBit}~\cite{Workgroup:2017bkh}), dark matter detection (with \textsf{DarkBit}~\cite{Cornell:2017opo}), flavour physics (with \textsf{FlavBit}~\cite{Workgroup:2017myk}) and direct searches from colliders (with \textsf{ColliderBit}~\cite{Balazs:2017moi}). The individual observables and likelihoods are calculated using the public package \textsf{GAMBIT-1.1.3}~\cite{Athron:2017ard}. The observables that we use draw on many other external software packages: \textsf{micrOMEGAs 3.6.9.2}~\cite{Belanger:2014vza}, \textsf{DDCalc 1.0.0}~\cite{Workgroup:2017lvb}, \textsf{FlexibleSUSY 1.5.13}~\cite{Athron:2014yba}, \textsf{gamLike 1.0.0}~\cite{Workgroup:2017lvb}, \textsf{GM2Calc 1.3.0}~\cite{Athron:2015rva}, \textsf{HiggsBounds 4.3.1}~\cite{Bechtle:2008jh}, \textsf{HiggsSignals 1.4}~\cite{Bechtle:2013xfa}, \textsf{SuperIso 3.6}~\cite{Mahmoudi:2007vz} and \textsf{SUSY-HIT 1.5}~\cite{Muhlleitner:2003vg}.
	
	\begin{figure}[ht]
		\includegraphics[width=8.5cm,height=12cm]{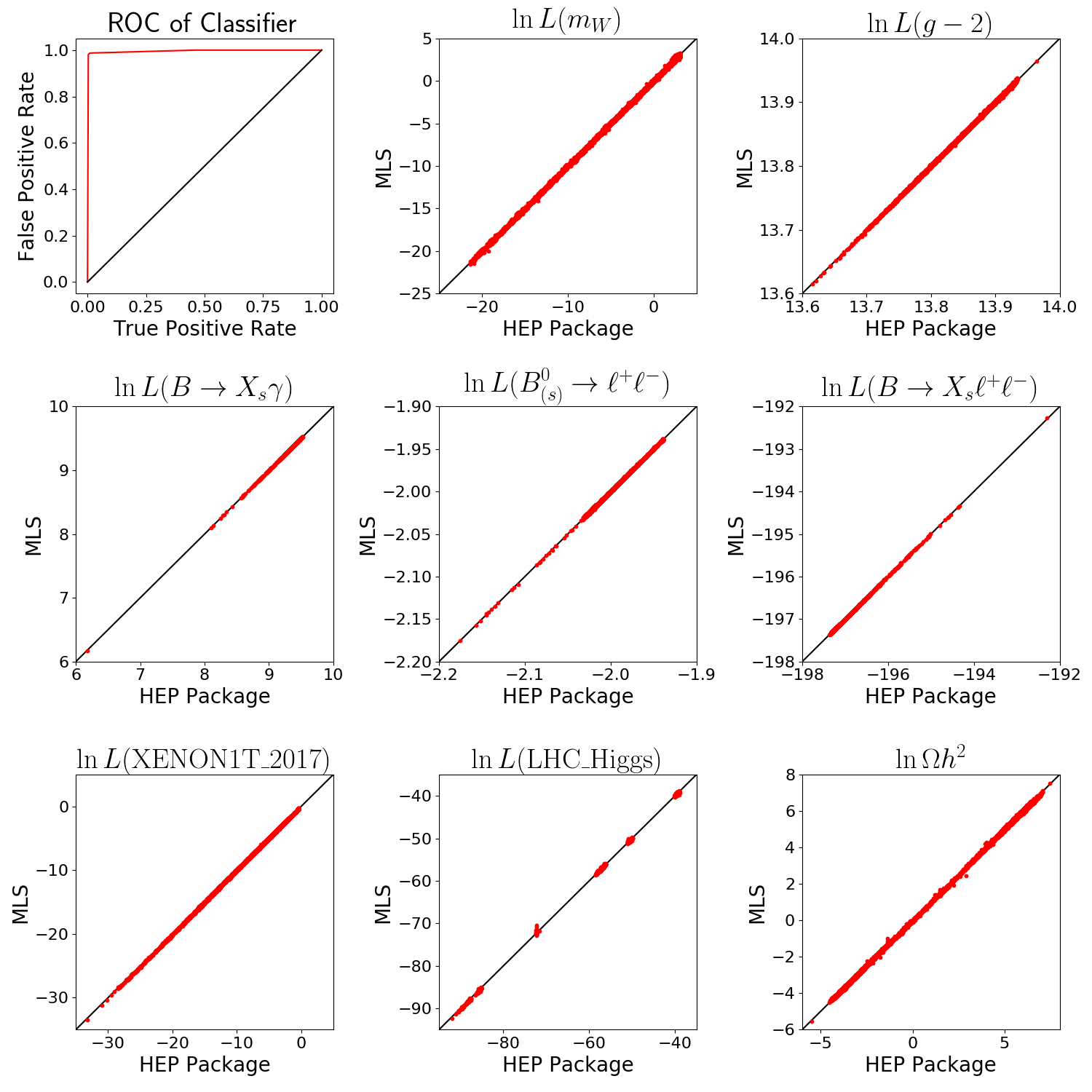}
		\caption{The ROC of MLS classifier for determining physical and unphysical parameter points of CMSSM (top-left panel). The relations between log-likelihoods of physical observables predicted by MLS regressor and the corresponding ones calculated by HEP packages (other panels).}
		\label{cmssm_fit}
	\end{figure}
	In Fig.~\ref{cmssm_fit}, we show the receiver operating characteristic curve (ROC) of MLS classifier for determining physical and unphysical parameter points of CMSSM on top-left panel. The sharpness of ROC demonstrates that MLS can efficiently recognize the physical and non-physical samples. Besides, in Fig.~\ref{cmssm_fit}, we present the relations between log-likelihoods of physical observable predicted by MLS regressor and the corresponding ones calculated by HEP packages. It can be seen that MLS regressor can predict the values of physical observables precisely, which are very close to their theoretical values.

	\begin{figure}[h]
		\includegraphics[width=8.5cm]{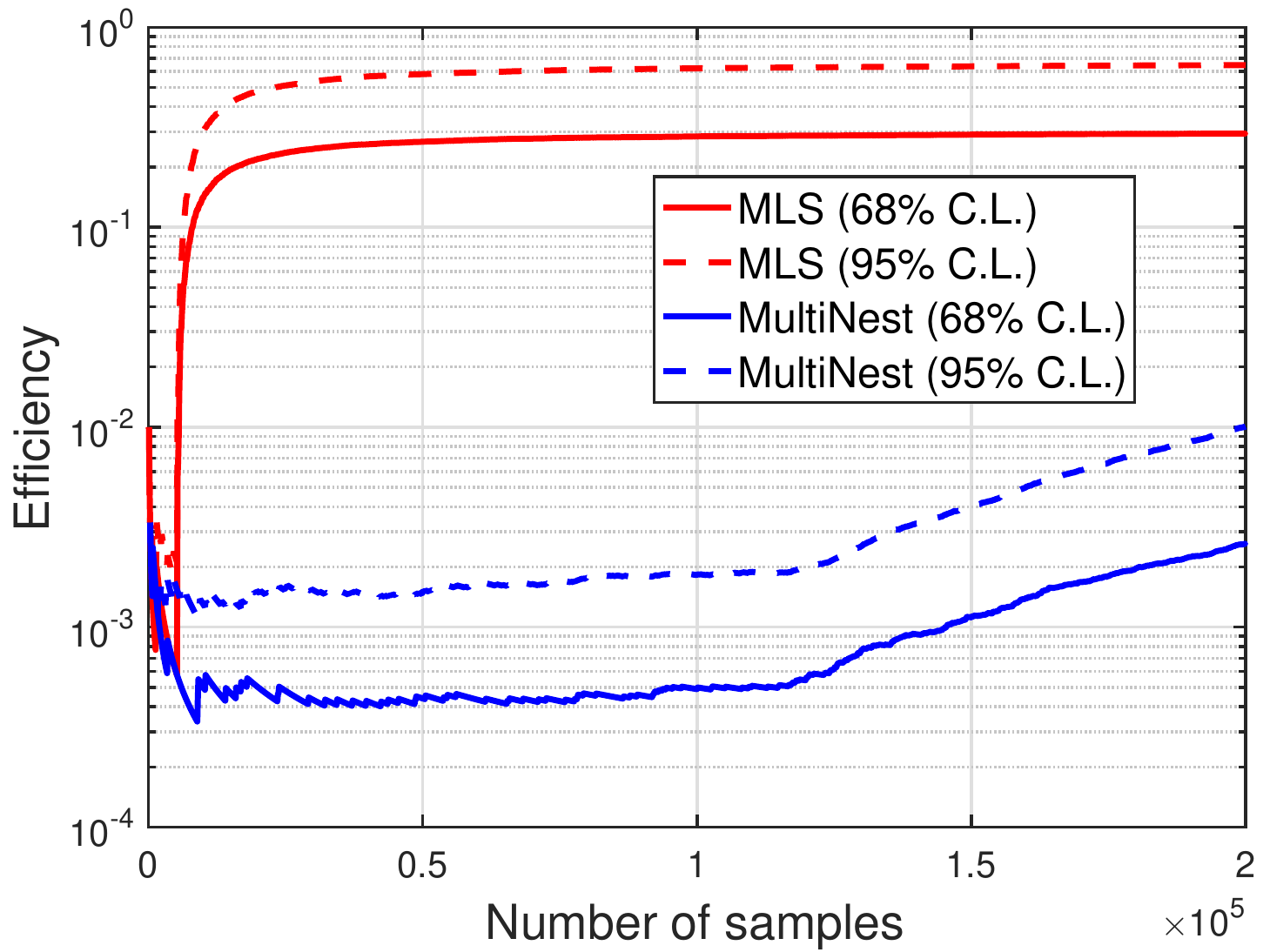}
		\caption{The efficiency of the MLS with 5,000 initial points and MultiNest for samples locating in 68\% (95\%) C.L. region of CMSSM parameter space. For both methods, the total number of samples are 200,000.}
		\label{cmssm_effi}
	\end{figure}
	
	With the above likelihood, we run MLS with 5,000 initial random samples. After evaluating 200,000 parameter points, we find that 58,719 (129,249) samples locate within 68\% (95\%) C.L. region of CMSSM parameter space. In fact, most of the survived regions can be discovered with about 20,000 samples. As a comparison, we perform a MultiNest scan with 5,000 live points, which is same as the setting in GAMBIT ~\cite{Athron:2017qdc}. We find that 523 (2016) out of 200,000 samples are within 68\% (95\%) C.L. region of CMSSM parameter space. The dependence of sampling efficiency on the number of samples is given in the Fig.~\ref{cmssm_effi}. If the number of samples increases enough, the efficiency of MultiNest may be comparable with the MLS.
	
	\begin{figure}[h]
		\includegraphics[width=8.5cm,height=7cm]{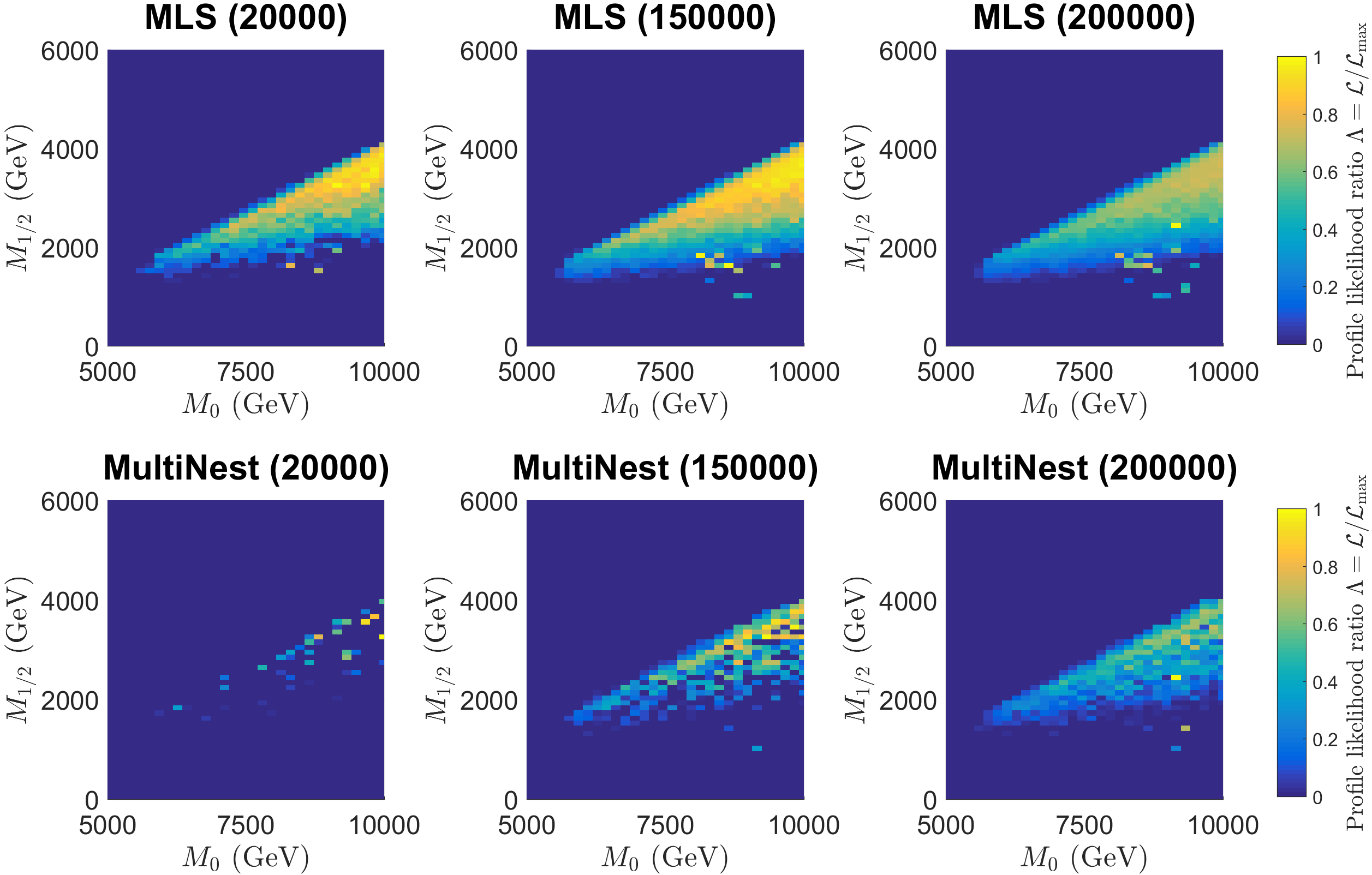}
		\caption{The profile likelihood ratio on the $M_0 - M_{1/2}$ plane for 20,000, 150,000 and 200,000 samples obtained from MLS (top panel) and MultiNest (bottom panel), respectively.}
		\label{cmssm_profile}
	\end{figure}
	In Fig.~\ref{cmssm_profile}, we show the profile likelihood ratio on the $M_0 - M_{1/2}$ plane for 20,000, 100,000, 150,000 and 200,000 samples obtained from MLS (top panel) and MultiNest (bottom panel), respectively. In such a region, the relic density of neutralino dark matter is saturated through the chargino co-annihilation or $A/H$ funnel. It can be seen that the chargino co-annihilation region (the large continuous region in Fig.~\ref{cmssm_profile} for the MLS scan is consistent with the MultiNest scan, as well as the result in GAMBIT~\cite{Athron:2017qdc}. As for the $A/H$ funnel region, i.e. the sporadic/fragmentary regions around $M_0\simeq 8~\mathrm{TeV}$ and $M_{1/2}\simeq2~\mathrm{TeV}$, the MLS method can collect more samples than MultiNest for 200,000 samples. So we can expect that the MLS can have a higher efficiency than the MultiNest in the case of limited computing resource.
	
	\textit{Model-4}.
Another popular SUSY model is the $R$-parity conserving MSSM, in which the lightest neutralino can be a natural WIMP dark matter. With the current data of LHC and DM experiments, the bino-like DM in the MSSM has to be heavier than 30 GeV when the lighter CP-even Higgs boson ($h$) is SM-like~\cite{Ambrogi:2017lov, Abdughani:2017dqs}. However, in the alignment limit~\cite{Carena:2013ooa}, the heavier CP-even scalar ($H$) can serve as the observed 125 GeV Higgs boson while the other scalar $h$ can be very light~\cite{Bechtle:2016kui, Profumo:2016zxo}. This provides a possibility that the bino-like DM below 30 GeV can saturate the DM relic density with the help of a light $h$.
	
We numerically find such an alignment in the MSSM using the MLS and MultiNest methods, respectively, under the following experimental constraints: (1) The SM Higgs boson is chosen as the heavier CP-even $H$; (2) All samples should be consistent with Higgs data, which is imposed by the \textsf{HiggsBounds-4.3.1}~\cite{Bechtle:2013wla} and \textsf{HiggsSignals 1.4.0}~\cite{Bechtle:2013xfa} as $\chi^2_\mathrm{HS}$; (3) DM relic density $\Omega h^2$ is within $3\sigma$ of the observed value~\cite{Ade:2015xua}. We scan the parameters as following:
	\begin{eqnarray}
	&& M_1 = 2 \sim 20~\mathrm{GeV}, \quad M_2 = M_3 = 2000~\mathrm{GeV} \nonumber \\
	&& \mu = 7000 \sim 9000~\mathrm{GeV}, \quad  \tan\beta = 2 \sim 10 \nonumber \\			
	&& m_{H^\pm} = 155~\mathrm{GeV}, \quad m_{\tilde{L}_3} = m_{\tilde{e}_3} = 1500~\mathrm{GeV} \nonumber \\
	&& m_{\tilde{Q}_{1,2}} = m_{\tilde{u}_{1,2}} = m_{\tilde{d}_{1,2}} = 2000~\mathrm{GeV} \nonumber \\
	&& m_{\tilde{Q}_3} = m_{\tilde{u}_3} = m_{\tilde{d}_3} = m_{\tilde{L}_{1,2}} = m_{\tilde{e}_{1,2}} = 1000~\mathrm{GeV} \nonumber \\
	&& A_t = A_b = A_\tau = -100~\mathrm{GeV}
	\end{eqnarray}
The Higgs mass and DM relic density are calculated with \textsf{FeynHiggs 2.13.0}~\cite{Heinemeyer:1998yj} and \textsf{micrOMEGAs 4.3.2}~\cite{Belanger:2010gh}, respectively.
	
	\begin{figure}[ht]
		\includegraphics[width=8.5cm]{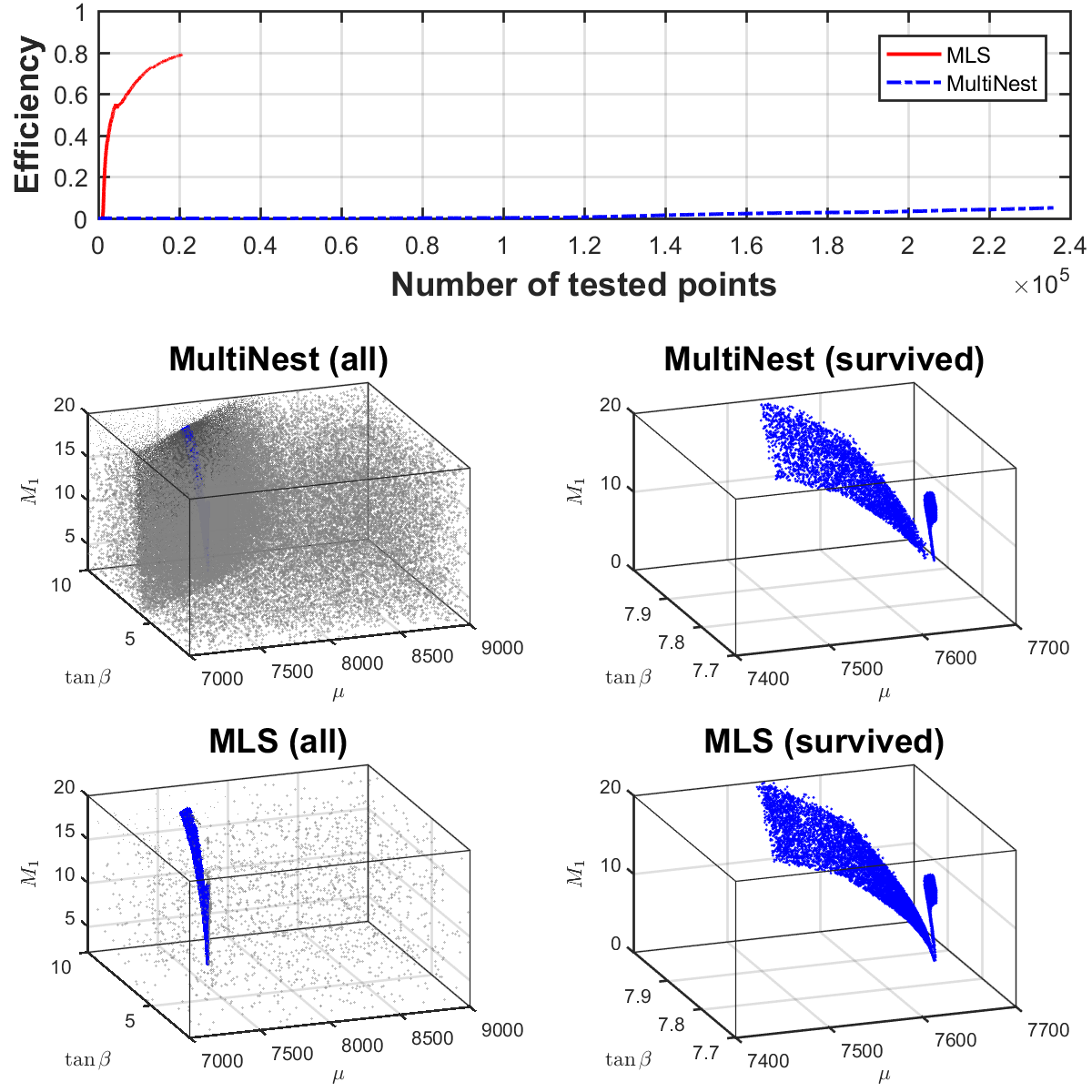}
		\caption{Same as Fig.~\ref{toy_eggbox_dist}, but for MSSM. The blue and gray points denote allowed and excluded (including non-physical) tested points, respectively.}
		\label{susy}
	\end{figure}
	
	To evaluate the parameter space, we construct the likelihood function as
	\begin{eqnarray}
	\label{lightDM}
	\mathcal{L} = \theta(3 - |m_H - 126|) \times \theta(112.7273 - \chi^2_\mathrm{HS}) \nonumber \\ \times \theta(0.03651 - |\Omega h^2 - 0.1186|) \times \exp\{-0.15 m_h\},
	\end{eqnarray}
	where $\theta(x)$ is the Heaviside step function and the last term ensures the light Higgs boson $h$ is as light as possible. We use 4 deep NN regressors to learn the $m_h$, $m_H$, $\chi^2_\mathrm{HS}$, $\ln\Omega h^2$, respectively, and 1 deep NN classifier to exclude non-physical points. Both have 4 hidden layers (each contains 60 neurons), except that the classifier has a sigmoid output neuron and trained with the binary cross-entropy loss function. The standard Adam optimizer is adopted to train them up to 2000 epochs. In the phase of recommendation, the candidates are first filtered by the classifier, and then only the remaining candidates are fed into the regressors to predict their corresponding likelihoods. The combined HEP packages averagely need 2 seconds to evaluate one parameter point, while the ML models can calculate $10^6$ parameter points in one second.
	
	In Fig.~\ref{susy}, we show the efficiency and coverage situation of the MLS and MultiNest to find the interested samples satisfying the requirements (1)-(4), which are in two sperate parameter regions. For the MLS, we use 1000 random samples to initialize the NNs and recommend 100 parameter points with 5 additional random ones in each iteration. It can be seen that the efficiency of MLS increases rapidly, which can reach 80\% when $2 \times 10^4$ samples are collected. On the other hand, the optimized MultiNest with 1000 live points has a low efficiency and has to run up to $2.35 \times 10^5$ tested points to discover the target regions. Even though, it has a poor coverage in the small $M_1$ region, as a comparison with our MLS.
	
	
	\textit{Conclusions.}
	Machine learning is becoming a powerful tool for the study of new physics, which can precisely find hidden patterns in complex models with least samples. We proposed a self-exploration method MLS to achieve a fast and reliable exploration of models with multi-parameter or equivalent solutions with a finite separation. We applied it to investigate the parameter space of MSSM with heavy Higgs $H$ being the SM-like Higgs boson in the alignment limit and subspace of the CMSSM as well. We find that MLS can significantly reduce the computational cost and ensure a better discovery of target regions, as a comparison with other conventional analysis methods.

	
	\textit{Acknowledgment.}
	We appreciate Yang Zhang very much for his great help of using GAMBIT. We also thank Ben Allanach, Farhan Feroz, Mike Hobson and Christopher Lester for helpful comments and discussions. This work was supported by the National Natural Science Foundation of China (NNSFC) under grant No. 11705093 and 11675242, by the CAS Center for Excellence in Particle Physics (CCEPP), by the CAS Key Research Program of Frontier Sciences and by a Key R\&D Program of Ministry of Science and Technology under number 2017YFA0402200-04.
	


%
	
\end{document}